\documentclass[twoside,12pt]{article}
\expandafter\let\csname equation*\endcsname\relax
\expandafter\let\csname endequation*\endcsname\relax
\usepackage{amsmath}
\usepackage{amssymb}
\usepackage{mathptmx}
\usepackage{graphicx}
\usepackage{cite}
\usepackage{color}
\usepackage{url}
\usepackage[unicode=true,
 bookmarks=false,
 breaklinks=false,pdfborder={0 0 1},backref=section,colorlinks=false]{hyperref}
\usepackage[latin9]{inputenc}
\usepackage{mathrsfs}

\usepackage{indentfirst}
\usepackage{bm}
\usepackage{epstopdf}

\graphicspath{{.}{./EPS/}{./Figures/}}

\renewcommand{\thesection}{\arabic{section}.\hspace{-.4cm}}
\renewcommand{\thesubsection}{\arabic{section}.\arabic{subsection}.\hspace{-.4cm}}
\renewcommand{\thesubsubsection}{\arabic{section}.\arabic{subsection}.\arabic{subsubsection}.\hspace{-.3cm}}

\begin{document}


\thispagestyle{empty} \vspace*{0.8cm}\hbox
to\textwidth{\vbox{\hfill\huge\sf Commun. Theor. Phys.\hfill}}
\par\noindent\rule[3mm]{\textwidth}{0.2pt}\hspace*{-\textwidth}\noindent
\rule[2.5mm]{\textwidth}{0.2pt}


\begin{center}
\LARGE\bf Measuring fine molecular structures with luminescence signal from
an alternating current scanning tunneling microscope
\end{center}

\footnotetext{\hspace*{-.45cm}\footnotesize $^\dag$Corresponding author: Hui Dong, Guohui Dong, E-mail: hdong@gscaep.ac.cn,20210076@sicnu.edu.cn}

\begin{center}
\rm Fei Wen$^{1}$, \ \ Guohui Dong$^{2\dagger}$ \ and  \ Hui Dong$^{1\dagger}$
\end{center}

\begin{center}
\begin{footnotesize} \sl
${}^{\rm 1}$ Graduate School of China Academy of Engineering Physics, Beijing
100084, China\\
${}^{\rm 2}$ School of Physics and Electronic Engineering, Sichuan Normal
University, Chengdu 610068, China\\
\end{footnotesize}
\end{center}

\begin{center}
\footnotesize (Received XXXX; revised manuscript received XXXX)

\end{center}

\vspace*{2mm}

\begin{center}
\begin{minipage}{15.5cm}
\parindent 20pt\footnotesize
In scanning tunneling microscopy induced luminescence (STML), the
photon counting is measured to reflect the single-molecule properties,
e.g., the first molecular excited state. The energy of the first excited
state is typically determined by a rising position of the photon counting
as a function of the bias voltage between the tip and the substrate.
It remains a challenge to determine the precise rise position of the
current due to the possible experimental noise. In this work, we propose
an alternating current version of STML to resolve the fine structures
in the photon counting measurement. The measured photon counting and
the current at the long-time limit show a sinusoidal oscillation.
The zero-frequency component of the current shows knee points at the
precise voltage as the fraction of the detuning between the molecular
gap and the DC component of bias voltage. We propose to measure the
energy level with discontinuity of the first derivative of such zero-frequency
component. The current method will extend the application of STML
in terms of measuring molecular properties. 
\end{minipage}
\end{center}

\begin{center}
\begin{minipage}{15.5cm}
\begin{minipage}[t]{2.3cm}{\bf Keywords:}\end{minipage}
\begin{minipage}[t]{13.1cm}
alternating current scanning tunneling
microscope, inelastic electron scattering, single-molecule electroluminescence,
molecular energy levels
\end{minipage}\par\vglue8pt

\end{minipage}
\end{center}

\section{Introduction}
Measuring the induced luminescence in scanning tunneling microscopy
(STML) is currently arising as a powerful tool to detect the single-molecular
properties, such as energy levels and optical responses \cite{Zhang2017}.
The technique of generating light from a metal-insulator-metal tunneling
junction was discovered by Lambe and McCarthy in 1976 \cite{Lambe_1976}.
Then the light emission with the nearly atomic spatial resolution
is reported for a scanning tunneling microscopy (STM) \cite{Coombs_1988}.

The origin of emitted light in STM had been controversial that whether
the photon is emitted from molecules. Intuitively, the transition
involving molecular states could lead to molecular luminescence as
the first origin. And the energy transfer from an excited molecular
state to the metal substrate may also contribute to light emission,
known as the quenching process as the second origin. Berndt et al.
reported spatially resolved photon emission from STM junction \cite{Berndt1993}.
And high emission efficiency from molecules was guaranteed by an oxide
layer which blocks such quenching process \cite{Flaxer1993}. To observe
the emission solely from molecules, a decoupling layer separated the
molecule from the substrate is needed \cite{Hoffmann_2002}. By virtue
of the decoupling proposal, molecular luminescence was realized with
different decoupling layers and various substrates. The fluorescence
from individual molecule is observed for porphyrin molecules adsorbed
on a thin aluminum oxide (Al$_{2}$O$_{3}$) film covered on a metal
NiAl(100) surface \cite{Qiu_2003,Qiu2004} and for molecules on the
organic film as the decoupling layer on the metallic substrate \cite{Dong_2004,Guo_2004,Guo2005,Guo_2005,Liu2007,Liu2007a,Liu2008,Ino2008}.
The ultra-thin insulating NaCl film was shown as a good decoupling
layer \cite{Guo2016} for the observation of luminescence, e.g., for
the individual pentacene \cite{Repp2005} and C$_{60}$ \cite{_avar_2005}
molecules from a metallic substrate . The advantage of strong enhanced
molecular fluorescence caused by the decoupling method \cite{Liu_2006,Liu_2009,Yan2007,You_2017}
makes the sandwich structure with metallic tip, decoupling layer and
the metallic substrate as a feasible platform for the STML experiments.
And our current work will focus on such structure.

In general, the theory of STML includes three mechanisms, i.e., the
inelastic electron scattering (IES) mechanism, the charge injection
(CI) mechanism and the gap plasmon mechanism. We focus on the IES
mechanism where the electron tunnels from one electrode to the other
inelastically while exciting the molecule in the gap. In the sandwich
setup, the tunneling current as well as the luminescence photon is
detected as a function of the bias voltage applied between the tip
and the substrate. Once the energy of the tunneling electron is above
a molecular transition gap, the single molecule can be pumped to an
excited state and then fluoresces. Thus a rise of the photon counting
can be observed at the position where the tunneling electron energy
matches the molecular transition energy \cite{Dong_2020}. And the
molecular energy level can be determined by the rise position of the
photon counting \cite{Du_2016}. Yet, it is difficult to accurately
determine the energy level due to the possible noise in the experiments
\cite{Imada_2021}.

In this paper, we propose an AC-STML method to detect the fine molecular
levels. Originally, STM with alternating current \cite{Guyon_2005}
was developed to probe the noise spectrum. Here, we extend its application
in molecular structure detection. To resolve the molecular structure,
we calculate the current and the luminescence photon counting for
the AC bias STM with perturbation theory and express the current in
the series of the Bessel's function. We find that the measured photon
counting and the inelastic current oscillate with time at the long-time
limit. The zero-frequency component of the Fourier transformation
of the current shows knee points at the precise voltage as the fraction
of the detuning between the molecular gap and the DC (time-independent)
component of bias voltage. The fine molecular structure can be determined
specifically with the knee points in the DC current (the zero-frequency
component of the AC current) as a function of the driving AC frequency.

The rest of the paper is organized as follows. In Sec. \ref{sec:Method},
we describe our model of STML with AC bias and calculate the inelastic
tunneling current in the time domain through the Bessel's function
of the first kind. Then we obtain the zero-frequency current in the
frequency domain. Section \ref{sec:Results} shows the current and
the first derivative of the current as a function of AC frequency.
We summarize the main contributions in Section \ref{sec:Conclusions}.

\section{Methods\label{sec:Method}}

\subparagraph*{The inelastic current in STML.}

\begin{figure}[t]
\begin{centering}
\includegraphics[width=8.5cm]{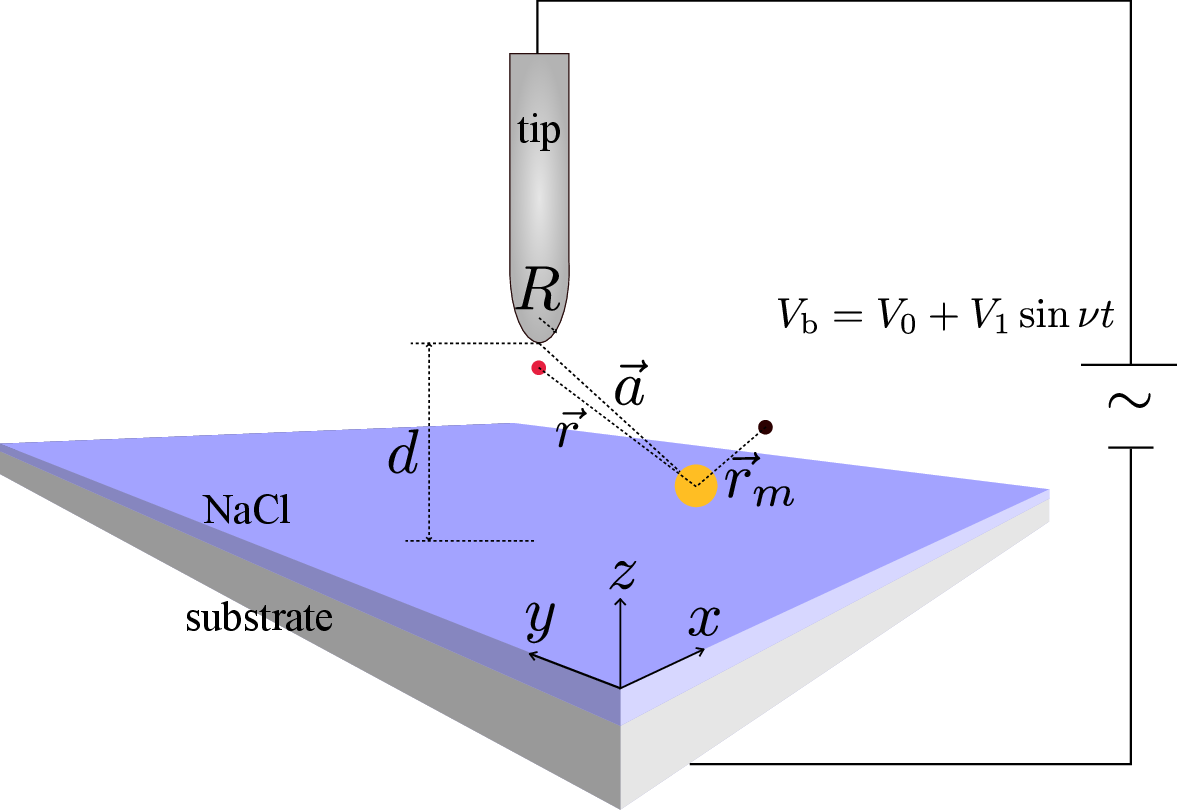} 
\par\end{centering}
\caption{\label{fig:The-schematic-diagram}The schematic diagram of the AC
STML system. In the dipole approximation, the yellow (black) sphere
represents the molecular nucleus (molecular electron). The molecule
is separated from the metallic substrate by the decoupling layer.
The red sphere stands for the tunneling electron. The radius of the
tip apex is $R$. $d$ represents the distance between the bottom
of the tip and the decoupling layer. $\vec{r}$ ($\vec{r}_{m}$, $\vec{a}$)
means the distance from the tunneling electron (molecular electron,
the center of the tip's apex) to the nucleus. The AC bias $V_{b}\left(t\right)=V_{0}+V_{1}\sin\left(\nu t\right)$is
applied between the tip and the substrate.}
\end{figure}

We sketch the setup of the AC-STML system in Fig. \ref{fig:The-schematic-diagram}.
The molecule is decoupled by a thin NaCl layer (the blue layer) from
the metal substrate (the grey layer). Here, we describe the molecule
with dipole approximation \cite{Dong_2020}. The nucleus is marked
by the yellow sphere and the molecular electron is marked by the black
sphere. The tip generates a tunneling electron (the red sphere) by
the AC bias applied between the tip and the substrate. The AC bias
$V_{b}\left(t\right)=V_{0}+V_{1}\sin\left(\nu t\right)$ contains
a time-independent part $V_{0}$ and a sinusoidal part with amplitude
$V_{1}$ and frequency $\nu$. The radius of the tip's apex is $R$,
and $d$ denotes the distance from the bottom of the tip to the substrate.
The molecular nucleus is set as the origin of the coordinate system.
$\vec{r}$, $\vec{r}_{m}$ and $\vec{a}$ denote the coordinates of
the tunneling electron, the molecular electron and the center of the
tip's apex, respectively. The tunneling electron interacts with the
molecule via the Coulomb interaction. The molecule can be excited
by the tunneling electron and subsequently emit photons via spontaneous
emission.

The Hamiltonian in our AC-STML system is divided into three parts,
\begin{align}
\hat{H}_{\mathrm{total}}=\hat{H_{\mathrm{el}}}+\hat{H}_{m}+\hat{H}_{\mathrm{el-m}},
\end{align}
where $\hat{H}_{\mathrm{el}}=-\hbar^{2}\nabla^{2}/2m_{e}+V\left(\vec{r}\right)$
corresponds to the Hamiltonian of the tunneling electron. $m_{e}$
is the electron mass, and $V\left(\vec{r}\right)$ is the potential
for the tunneling electron at the position $\vec{r}$. $\hat{H}_{m}=E_{g}\left|\chi_{g}\right\rangle \left\langle \chi_{g}\right|+E_{e}\left|\chi_{e}\right\rangle \left\langle \chi_{e}\right|$
is the Hamiltonian of the molecule which is simplified as a two-level
system \cite{Nian_2018,Zhang_2016,Chen_2019}. $\left|\chi_{g}\right\rangle $
is the molecular ground state with energy $E_{g}$, and $\left|\chi_{e}\right\rangle $
is the molecular excited state with energy $E_{e}$. $\hat{H}_{\mathrm{el-m}}\simeq-e\hat{\vec{\mu}}\cdot\vec{r}/\left|\vec{r}\right|^{3}$
is the Coulomb interaction between the tunneling electron and the
molecule. $e$ is the electron charge, and $\hat{\vec{\mu}}$ is the
electric dipole moment operator of the molecule \cite{Dong_2020}.
This interaction will induce the molecular transition between its
two states in the inelastic tunneling process.

The electron wavefunctions in the state $k$ of the tip and in the
state $n$ of the substrate are written \cite{Bardeen_1961} as 
\begin{align}
\left\langle \vec{r}|\phi_{k}\right\rangle  & =A_{k}\frac{\mathrm{e}^{-\kappa_{k}\left(\left|\vec{r}-\vec{a}\right|-R\right)}}{\kappa_{k}\left|\vec{r}-\vec{a}\right|},\label{eq:tip-function}\\
\left\langle \vec{r}|\varphi_{n}\right\rangle  & =B_{n}\mathrm{e}^{-\kappa_{n}\left|z\right|},
\end{align}
where $\kappa_{k}=\sqrt{-2m_{e}\xi_{k}}/\hbar$ ($\kappa_{n}=\sqrt{-2m_{e}E_{n}}/\hbar$)
characterizes the decay factor of the tunneling electron with energy
$\xi_{k}$ ($E_{n}$) in the tip (substrate). $A_{k}$ and $B_{n}$
are the normalized coefficients derived from the first-principle calculation
\cite{Tersoff_1983,Tersoff_1985}.

At a given bias, both the elastic and the inelastic tunneling occur.
The interaction between the molecule and the tunneling electron is
small and is treated as a perturbation. Thus the probability of the
molecule in its excited state is quite low \cite{Chong2021}. At the
beginning, we assume that the molecule is in its ground state and
the tunneling electron is in the state $n$ of the substrate, i.e.,
$\left|\Psi\left(0\right)\right\rangle =\left|\chi_{g}\right\rangle \left|\varphi_{n}\right\rangle $
noticing the typical low temperature in the STML experiments \cite{Repp2005,Qiu_2003,Qiu2004}.
Using the time-dependent perturbation theory, we find that at time
$t$, the system evolves to state 
\begin{align}
\left|\Psi\left(t\right)\right\rangle  & =\mathrm{e}^{-\mathrm{i}\left(E_{n}+E_{g}\right)t/\hbar}\left|\chi_{g}\right\rangle \left|\varphi_{n}\right\rangle \nonumber \\
 & +\sum_{k}c_{g,k}\left(t\right)\left|\chi_{g}\right\rangle \left|\phi_{k}\right\rangle +\sum_{k}c_{e,k}\left(t\right)\left|\chi_{e}\right\rangle \left|\phi_{k}\right\rangle ,
\end{align}
where $c_{g,k}\left(t\right)$ is the elastic tunneling amplitude
and $c_{e,k}\left(t\right)$ is the inelastic tunneling amplitude.
The dynamical equations of the tunneling amplitudes read 
\begin{align}
\mathrm{i}\hbar\frac{\mathrm{d}c_{g,k}\left(t\right)}{\mathrm{d}t} & =\left(\tilde{\xi}_{k}\left(t\right)+E_{g}\right)c_{g,k}\left(t\right)+\mathrm{e}^{-\mathrm{i}\left(\tilde{E}_{n}+E_{g}\right)\frac{t}{\hbar}}\mathcal{M}_{n,k},\\
\mathrm{i}\hbar\frac{\mathrm{d}c_{e,k}\left(t\right)}{\mathrm{d}t} & =\left(\tilde{\xi}_{k}\left(t\right)+E_{e}\right)c_{e,k}\left(t\right)\nonumber \\
 & +\mathcal{N}_{s,t}|_{V_{b},E_{n}\rightarrow\xi_{k}}\mathrm{e}^{-\mathrm{i}\left(\tilde{E}_{n}+E_{g}\right)\frac{t}{\hbar}},
\end{align}
where $\mathcal{M}_{n,k}$ is the transition matrix element in the
elastic tunneling and $\mathcal{N}_{s,t}\mid_{V_{b},E_{n}\rightarrow\xi_{k}}$
is the transition matrix element from the substrate's state to the
tip's state under the AC bias $V_{b}$ \cite{Dong_2020}. $\tilde{E}_{n}$
($\tilde{\xi}_{k}$) is the electron energy in the substrate (tip)
under the bias. In this work, we apply an AC bias $V_{b}\left(t\right)$
between the electrodes. In this case, we have the simple relations,
$\widetilde{\xi}_{k}=\xi_{k}+eV_{0}+eV_{1}\sin\left(\nu t\right)$
and $\widetilde{E}_{n}=E_{n}$.

The inelastic tunneling rate is written as 
\begin{align}
\frac{\mathrm{d}}{\mathrm{d}t}\left|c_{e,k}\left(t\right)\right|^{2} & =\frac{2}{\hbar^{2}}\mathcal{N}_{s,t}^{2}|_{V_{b},E_{n}\rightarrow\xi_{k}}\mathrm{Re}\left\{ \mathrm{e}^{-\mathrm{i}\left[Bt-\frac{eV_{1}}{\hbar\nu}\cos\left(\nu t\right)\right]}\right.\nonumber \\
 & \left.\times\int_{0}^{t}\mathrm{d}\tau\mathrm{e}^{\mathrm{i}\left[B\tau-\frac{eV_{1}}{\hbar\nu}\cos\left(\nu\tau\right)\right]}\right\} ,
\end{align}
where we have used the notation $B\equiv\left(\xi_{k}-E_{n}+E_{\mathrm{eg}}+eV_{0}\right)/\hbar$.
The inelastic electron current is 
\begin{align}
I_{s,t}\left(t\right) & =e\sum_{n,k}F_{\mu_{0},T}\left(E_{n}\right)\left[1-F_{\mu_{0},T}\left(\xi_{k}\right)\right]\frac{\mathrm{d}}{\mathrm{d}t}\left|c_{e,k}\left(t\right)\right|^{2},
\end{align}
where $F_{\mu_{0},T}\left(E\right)=1/\left\{ \exp\left[\left(E-\mu_{0}\right)/k_{\mathrm{B}}T\right]+1\right\} $
is the Fermi-Dirac distribution of the electron in the tip or substrate
at energy $E$ and temperature $T$. $k_{\mathrm{B}}$ is the Boltzmann
constant and $\mu_{0}$ is the Fermi energy. The subscript of $I_{s,t}$
means that the tunneling electron passes from the substrate to the
tip. In experiments, STML is performed under a very low temperature,
e.g., 5K or 13K \cite{Repp2005,Qiu_2003,Qiu2004}. Therefore, the
Fermi-Dirac distribution is simplified as $F_{\mu_{0},T}\left(E\right)=1$
for $E<\mu_{0}$ and $F_{\mu_{0},T}\left(E\right)=0$ for $E\geq\mu_{0}$.
The inelastic current is thus rewritten as $I_{s,t}\left(t\right)=e\sum_{n,k}\frac{\mathrm{d}}{\mathrm{d}t}\left|c_{e,k}\left(t\right)\right|^{2}$.

By replacing the summation with integral and using Jacobi-Anger expansion
$\mathrm{e}^{\mathrm{i}a\cos\left(\nu t\right)}=\sum_{l=-\infty}^{\infty}\mathrm{e}^{\mathrm{i}l\pi/2}J_{l}\left(a\right)\mathrm{e}^{\mathrm{i}l\nu t}$,
we obtain the inelastic tunneling current explicitly as 
\begin{align}
I_{s,t}\left(t\right) & =\int_{-\infty}^{\mu_{0}}\mathrm{d}E_{n}\int_{\mu_{0}}^{0}\mathrm{d}\xi_{k}\rho_{s}\left(E_{n}\right)\rho_{t}\left(\xi_{k}\right)\frac{\mathrm{d}}{\mathrm{d}t}\left|c_{e,k}\left(t\right)\right|^{2}\nonumber \\
 & =\frac{2}{\hbar^{2}}\int_{2\mu_{0}}^{\mu_{0}}\mathrm{d}E_{n}\int_{\mu_{0}}^{0}\mathrm{d}\xi_{k}\rho_{s}\left(E_{n}\right)\rho_{t}\left(\xi_{k}\right)\nonumber \\
 & \times\mathcal{N}_{s,t}^{2}|_{V_{b},E_{n}\rightarrow\xi_{k}}\sum_{l,l'=-\infty}^{\infty}\left(-1\right)^{l'}\nonumber \\
 & \times J_{l}\left(\frac{eV_{1}}{\hbar\nu}\right)J_{l'}\left(\frac{eV_{1}}{\hbar\nu}\right)\left(B-l'\nu\right)^{-1}\nonumber \\
 & \times\{\cos\left[\left(l-l'\right)\nu t+\left(l+l'-1\right)\pi/2\right]\nonumber \\
 & -\cos\left[-\left(B-l\nu\right)t+\left(l+l'-1\right)\pi/2\right]\},\label{eq:time-depended current}
\end{align}
where $\rho_{s}\left(E\right)$ ($\rho_{t}$$\left(E\right)$) is
the density of state in the substrate (tip) at energy $E$. $E_{\mathrm{eg}}=E_{e}-E_{g}$
is the molecular energy gap, and $J_{i}\left(x\right)$ is the $i$-th
Bessel function of the first kind. In the derivation above, we have
used the property of the Bessel functions $J_{i}\left(-a\right)=\left(-1\right)^{i}J_{i}\left(a\right)$.
We change the range of the integral about $E_{n}$ from $\left[-\infty,\mu_{0}\right]$
to $\left[2\mu_{0},\mu_{0}\right]$, since most electrons of the metal
occupy states near the Fermi energy. The detailed derivations are
presented in Appendix A.

\subparagraph*{AC-current in frequency domain\label{sec:ac-current}.}

Different from DC voltage case, the system under the AC bias will
not reach a steady state with a constant current at the long-time
limit. Instead, the current oscillates with various Fourier frequencies.
The information of the energy level can be extracted from these components
in the Fourier transformations of $I_{s,t}\left(t\right)$ as following
\begin{align}
I_{s,t}\left(\omega\right) & =\int_{-\infty}^{+\infty}\mathrm{d}t\mathrm{e}^{\mathrm{i}\omega t}I_{s,t}\left(t\right)\nonumber \\
 & =\frac{2\pi}{\hbar^{2}}\int_{2\mu_{0}}^{\mu_{0}}\mathrm{d}E_{n}\int_{\mu_{0}}^{0}\mathrm{d}\xi_{k}\rho_{s}\left(E_{n}\right)\rho_{t}\left(\xi_{k}\right)\nonumber \\
 & \times\mathcal{N}_{s,t}^{2}|_{V_{b},E_{n}\rightarrow\xi_{k}}\sum_{l,l'=-\infty}^{\infty}\left(-1\right)^{l'}\nonumber \\
 & \times J_{l}\left(\frac{eV_{1}}{\hbar\nu}\right)J_{l'}\left(\frac{eV_{1}}{\hbar\nu}\right)\left(B-l'\nu\right)^{-1}\nonumber \\
 & \times\{\delta\left[\left(l-l'\right)\nu+\omega\right]\mathrm{e}^{\mathrm{i}\left(l+l'-1\right)\pi/2}\nonumber \\
 & +\delta\left[\left(l-l'\right)\nu-\omega\right]\mathrm{e}^{-\mathrm{i}\left(l+l'-1\right)\pi/2}\nonumber \\
 & -\delta\left[\left(B-l\nu\right)+\omega\right]\mathrm{e}^{-\mathrm{i}\left(l+l'-1\right)\pi/2}\nonumber \\
 & -\delta\left[\left(B-l\nu\right)-\omega\right]\mathrm{e}^{\mathrm{i}\left(l+l'-1\right)\pi/2}\}.\label{eq:fourier}
\end{align}
Here, we have used the relation $\int_{-\infty}^{+\infty}\mathrm{d}t\mathrm{e}^{\mathrm{i}\Omega t}=2\pi\delta\left(\Omega\right)$.

The photon counting in the AC system heavily depends on the AC inelastic
tunneling current we calculated above. As the molecule is firstly
excited to its excited state by the AC bias and then decays back through
the spontaneous emission process, the population equation of the molecule
excited state reads \cite{Dong_2020} $\dot{P}_{e}\left(t\right)=-\gamma P_{e}\left(t\right)+I_{s,t}\left(t\right)$,
where $P_{e}\left(t\right)$ stands for the excited-state population
and $\gamma$ is the spontaneous decay rate. Since the AC inelastic
tunneling current oscillates with various Fourier frequencies, we
divide the excited-state population $P_{e}\left(t\right)$ with respect
to the Fourier frequencies, i.e., $-\mathrm{i}\omega P_{e}\left(\omega\right)=-\gamma P_{e}\left(\omega\right)+I_{s,t}\left(\omega\right)$
where $P_{e}\left(\omega\right)=\int_{-\infty}^{\infty}P_{e}\left(t\right)\mathrm{e}^{\mathrm{i}\omega t}\mathrm{d}t$.
Thus, the zero-component current contributes to a steady photon counting
in the long-time limit while every nonzero-frequency component gives
one time-dependent photon counting with its corresponding frequency.

To find the energy levels, we consider the zero-frequency component
of the inelastic current (Eq. (\ref{eq:fourier})), i.e., $\omega=0$
as

\begin{align}
I_{s,t}\left(\omega=0\right) & =\frac{4\pi}{\hbar^{2}}\int_{2E_{f}}^{E_{f}}\mathrm{d}E_{n}\int_{E_{f}}^{0}\mathrm{d}\xi_{k}\rho_{s}\left(E_{n}\right)\rho_{t}\left(\xi_{k}\right)\nonumber \\
 & \times\mathcal{N}_{s,t}^{2}|_{V_{b},E_{n}\rightarrow\xi_{k}}\sum_{l,l'=-\infty}^{\infty}\left(-1\right)^{l'}\nonumber \\
 & \times J_{l}\left(\frac{eV_{1}}{\hbar\nu}\right)J_{l'}\left(\frac{eV_{1}}{\hbar\nu}\right)\left(B-l'\nu\right)^{-1}\nonumber \\
 & \times\cos\left[\left(l+l'-1\right)\pi/2\right]\nonumber \\
 & \times\left\{ \delta\left[\left(l-l'\right)\nu\right]-\delta\left(B-l\nu\right)\right\} .\label{eq:zero-fre component}
\end{align}
The result shown in Eq. (\ref{eq:zero-fre component}) retains the
current for the case with DC voltage ($V_{1}=0$) obtained in \cite{Dong_2020}
by noticing $J_{0}(0)=1$ and $J_{l}(0)=0$ for any $l\neq0$. In
the paper \cite{Dong_2020}, the current caused by the DC bias keep
constant at the long-time limit, while in our AC case, the inelastic
current oscillates with the time and cannot reach a steady state.
So we make Fourier transformation of the AC-induced current and investigate
the zero-frequency component of the current. In the long-time limit,
the time-independent photon counting is proportional to the zero-component
current, i.e., Eq. (\ref{eq:zero-fre component}).

\section{Results\label{sec:Results}}

To reveal the resonant conditions in the above current, we perform
the numerical calculation of tunneling current with parameters extracted
from the experimental setup. In the STML experiments, the metal used
for the tip and the substrate is typically chosen as gold (Au) \cite{Chong_2016,Gro_e_2017,Merino_2018,Doppagne_2018,Ros_awska_2018},
silver (Ag) \cite{Zhang_2013,Zhang_2016,Imada_2016,Imada_2017,Zhang2017,Zhang2017a,Chen_2019,Kimura_2019,Yang_2020,Zoh_2021}
and copper (Cu) \cite{Repp2005,Guo2005,Schneider_2012,Edelmann_2020}.
In our simulation of the STML current, the tip and the substrate are
made of silver with Fermi energy $\mu_{0}=-4.64\mathrm{eV}$. The
calculation of the current requires the Ag's density of state, which
is obtained from the book \cite{Papaconstantopoulos_2015} by the
spline interpolation of the discrete data points. The detail of the
obtained density of state was presented in our previous publications
\cite{Dong_2020,Dong_2021}. In the experiments, the molecular gap
is typically chosen around $1.5\mathrm{eV}\sim4\mathrm{eV}$ \cite{Meng_2015,Imada_2017,Zhang_2016,Doppagne_2017,Doppagne_2018,Zhang2017,Zhang2017a,Dole_al_2019}to
avoid the possible damage caused by the strong static electric field
between the tip and the substrate. For example, the energy gap between
the first singlet excited state and the ground state of the free-base
phthalocyanine (H$_{2}$Pc) molecule is $1.81\mathrm{eV}$\cite{Meng_2015,Imada_2017},
and the $Q\left(0,0\right)$ transition energy of zinc-phthalocyanine
(ZnPc) molecule is $1.90\mathrm{eV}$ \cite{Zhang_2016,Doppagne_2018,Doppagne_2017,Zhang2017a,Zhang2017,Dole_al_2019}.
Here, we choose that the molecular gap is $E_{\mathrm{eg}}=2.0\mathrm{eV}$.
In the scanning process, the distance $d$ between the tip and the
substrate is typically around several nanometers. And we have used
$d=0.5\mathrm{nm}$ and the radius of the tip apex is $R=0.5\mathrm{nm}$.

\subparagraph*{Tunneling current for the tip position $\vec{a}=\left(0,0,d\right)$.}

\begin{figure}[htb]
\begin{centering}
\includegraphics[width=8.5cm]{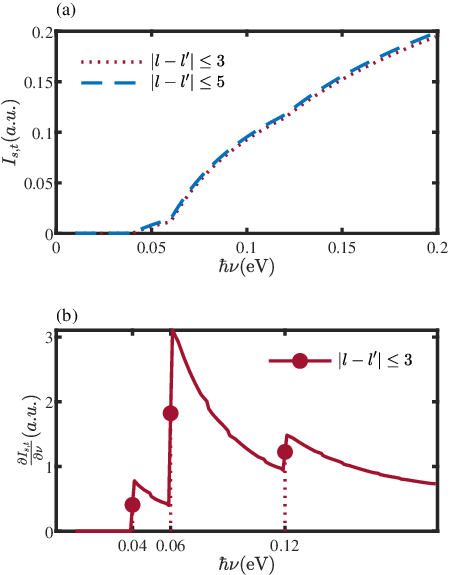} 
\par\end{centering}
\caption{\label{fig:convergency =00003D00003D00003D00003D000026 derivative_1}
(a) The convergency of the zero-frequency inelastic current with $\vec{a}=\left(0,0,d\right)$.
The red dotted (blue dashed) curve shows the current under the summation
of $\left|l\right|\protect\leq3$ and $\left|l-l'\right|\protect\leq3$
($\left|l-l'\right|\protect\leq5$). Both lines show the non-analyticity
of the current with respect to the oscillating frequency. (b) The
first derivative of the zero-frequency inelastic current with respect
to the oscillating frequency $\nu$. The tip is placed right above
the molecule. The red dots show the discontinuous data in the current,
which correspond to $\hbar\nu=0.04,\;0.06,$ and $0.12\mathrm{eV}$.
We have chosen the parameters as $l\in[-3,3],\;R=0.5\mathrm{nm},\;d=0.5\mathrm{nm},\;\vec{a}=\left(0,0,d\right),\;E_{\mathrm{eg}}=2\mathrm{eV},\;\mu_{0}=-4.64\mathrm{eV}$,
and $eV_{0}=-1.88\mathrm{eV}$. The ratio of the time-dependent voltage
amplitude over the oscillating frequency is fixed at $eV_{1}/\hbar\nu=2$.}
\end{figure}

Since the absolute value of the Bessel function decreases as its order
increases, it is reasonable to cut off the high order term of the
Bessel function of the summation in Eq. (\ref{eq:zero-fre component}).
Noticing that the factor $\cos\left[\left(l+l'-1\right)\pi/2\right]$
vanishes for the case where $l+l'$ is even, we consider the cutoffs
with $\left|l-l'\right|\leq3$ and $\left|l-l'\right|\leq5$ to check
the convergence of the current in Eq. (\ref{eq:zero-fre component}).
Fig. \ref{fig:convergency =00003D00003D00003D00003D000026 derivative_1}
(a) shows the zero-frequency inelastic tunneling current as a function
of AC frequency $\nu$. The parameters in the simulation are given
as follows, $l\in\left[-3,3\right],\:R=0.5\mathrm{nm},\;d=0.5\mathrm{nm},\;\vec{a}=\left(0,0,d\right),\;E_{\mathrm{eg}}=2\mathrm{eV},\;eV_{1}/\hbar\nu=2,\;\mu_{0}=-4.64\mathrm{eV}$
and $eV_{0}=-1.88\mathrm{eV}$. The tip is placed right above the
molecule. In Fig. \ref{fig:convergency =00003D00003D00003D00003D000026 derivative_1}
(a), the red dotted line reveals the current including the summation
of $\left|l\right|\leq3$ and $\left|l-l'\right|\leq3$, and the black
line shows that of$\left|l\right|\leq3$ and $\left|l-l'\right|\leq5$.
The coincidence of two curves demonstrates that the zero-frequency
inelastic current already converges with $\left|l-l'\right|=3$ and
$\left|l\right|\leq3$. Therefore, we use the cutoff $\left|l\right|\leq3$
and $\left|l-l'\right|\leq3$ in the following calculation.

The curve in Fig. \ref{fig:convergency =00003D00003D00003D00003D000026 derivative_1}
(a) also shows the discrete knee points. We numerically calculate
the first derivative of the inelastic tunneling current $I_{s,t}\left(\omega=0\right)$
with respect to the AC frequency $\nu$ and plot the result in Fig.
\ref{fig:convergency =00003D00003D00003D00003D000026 derivative_1}
(b). In Fig. \ref{fig:convergency =00003D00003D00003D00003D000026 derivative_1}
(b), the line reveals the discontinuity of the first derivative of
the zero-frequency current with the discontinuous spots located at
$\hbar\nu=0.04\mathrm{eV},\;0.06\mathrm{eV},\;\mathrm{and}\;0.12$eV.
By defining the detuning $\Delta=E_{\mathrm{eg}}+eV_{0}$, we find
that the discontinuous spots are in agree with the condition $\Delta-l'\hbar\nu=0$
with $l'=3,2,1$, respectively. Mathematically, such discontinuous
behavior can be understood with Jacobi-Anger expansion. The STML system
with AC bias $V_{0}+V_{1}\sin\left(\nu t\right)$ is equivalent to
the system with a series of DC bias $V_{l}^{\mathrm{eff}}=V_{0}-l'\hbar\nu/e$
($l'=0,\pm1,\pm2,...$). Once one effective bias $V_{l}^{\mathrm{eff}}$
matches the molecular energy gap, namely $V_{l}^{\mathrm{eff}}=E_{\mathrm{eg}}$,
a new contribution to the molecular excitation rate (the inelastic
current) emerges and results in one discontinuous point in its first
derivative curve. Therefore we have demonstrated that the discontinuous
spots reveal the fine detail of the molecule.

\begin{figure}[htb]
\begin{centering}
\includegraphics[width=8.5cm]{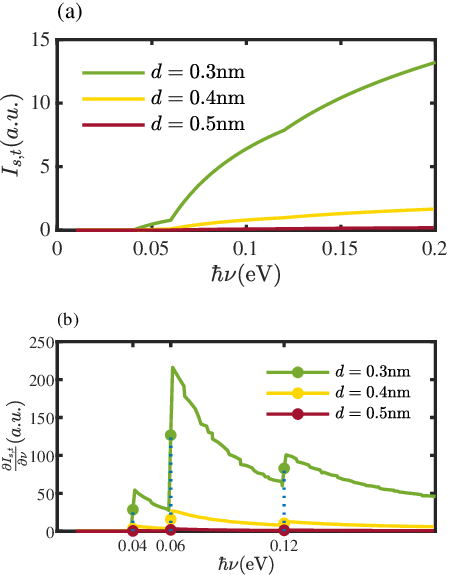}
\par\end{centering}
\caption{\label{fig:different_d}(a) The zero-frequency component of the inelastic
current with different height of the tip. The green, yellow and red
curves correspond to the inelastic current with $d=0.3\mathrm{nm}$,
$d=0.4\mathrm{nm}$ and $d=0.5\mathrm{nm}$ respectively. (b) The
first derivative of the zero-frequency inelastic current with respect
to the frequency $\nu$. The tip is placed right above the molecule.
The dotted markers in (b) show the discontinuous data at $\hbar\nu=0.04\mathrm{eV},\;0.06\mathrm{eV},$
and $0.12\mathrm{eV}$. The other parameters are the same as before.}
\end{figure}

To investigate the influence caused by the height of the tip, we choose
the parameter $d=0.3\mathrm{nm}$, $d=0.4\mathrm{nm}$ and $d=0.5\mathrm{nm}$
respectively. Fig. \ref{fig:different_d} (a) shows the zero-frequency
component of the inelastic current which is depend on the position
of the tip. The green, yellow and red curves correspond to the inelastic
current with the tip fixed at $d=0.3\mathrm{nm}$,$d=0.4\mathrm{nm}$
and $d=0.5\mathrm{nm}$ respectively. The inelastic current becomes
larger as the tip moves to the molecule. Because we modeled the electron
wavefunction of the tip as Eq.( \ref{eq:tip-function}), which decays
with the tip radius exponentially. When the STM tip approaches the
substrate, the electron wavefunction at the position of the tunneling
electron increases, resulting in the increase of the transition matrix
element. Finally, the inelastic current increases. Fig. \ref{fig:different_d}
(b) reveals the first derivative of the inelastic current with respect
to the frequency. The discontinuous data are marked by the dot. All
the curves have the discontinuous data at the same spots $\hbar\nu=0.04\mathrm{eV},\;0.06\mathrm{eV},$
and $0.12\mathrm{eV}$, which are in agree with the condition $\Delta-l'\hbar\nu=0$.
Although the inelastic current changes with the tip approaching the
molecule, the knee points of the current satisfy the resonant condition
which is independent on the position of the tip.

The advantage of our AC-STML is that the frequency can be tuned with
precision. In the DC bias case, the inelastic tunneling current has
a sudden rise from zero when the absolute value of the DC bias equals
to a critical quantity, i.e., the absolute value of the molecular
energy gap divided by the electron charge \cite{Dong_2020}. And one
can extract the information of the molecular energy gap from this
curve theoretically. However, due to the noise of the experiment,
the inelastic tunneling current curve changes smoothly near the critical
quantity. Thus one can only read out the energy gap roughly. In the
AC STML method, the point of the energy gap is featured as the knee
point of the zero-frequency component of the AC current. These points
show discontinuous and non-analytical property at the non-zero point
of the current curve. By numerically calculating the first derivative
of the zero-frequency component of the inelastic tunneling current
with respect to the AC frequency, the discontinuous points, i.e.,
the sudden rising points, correspond to the points of energy resonance.
We can read out discontinuous points directly and then obtain the
energy gap of the molecule.

The method of realizing our proposal includes two steps. Firstly,
the molecule is probed via DC bias. The molecular energy gap can be
roughly obtained through the rising point in the photon-emission spectrum.
And we estimate the rough value as $V_{0}$. Secondly, we add a nonzero
AC component to this DC voltage and apply this time-dependent bias
to the molecule. A series of knee points can be shown in the figure
of the zero-frequency inelastic tunneling current $I_{s,t}\left(\omega=0\right)$
as a function of AC frequency. Then, we can precisely determine these
non-analytical points through the first derivative of the inelastic
tunneling current $I_{s,t}\left(\omega=0\right)$ with respect to
the AC frequency. The precise molecular gap is given with the relation
$E_{\mathrm{eg}}-e\left|V_{0}\right|=l'\hbar\nu\left(l'=0,\pm1,\pm2,\pm3\right)$.

\subparagraph*{Tunneling current for the tip position $\vec{a}=\left(0.2\mathrm{nm},0,d\right)$.}

\begin{figure}[t]
\begin{centering}
\includegraphics[width=8.5cm]{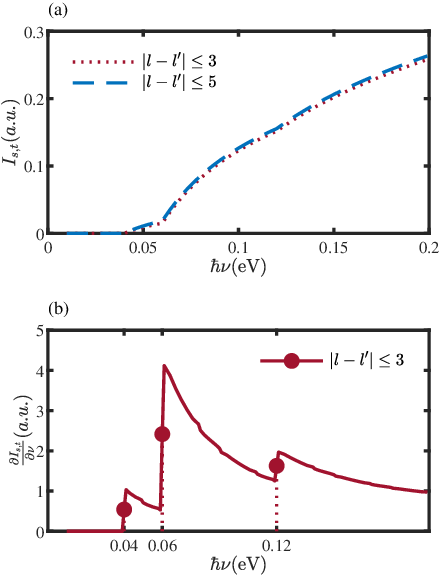}
\par\end{centering}
\caption{\label{fig:lateral_displacement}(a) The convergency of the zero-frequency
inelastic current in $\vec{a}=\left(0.2\mathrm{nm},0,d\right)$. The
red dotted line and the blue dashed line correspond to the condition
$\left|l-l'\right|\protect\leq3$ and $\left|l-l'\right|\protect\leq5$,
respectively. (b) The first-order derivative of the zero-frequency
inelastic current about the bias frequency $\nu$ in $\vec{a}=\left(0.2\mathrm{nm},0,d\right)$.
The other parameters are the same as before.}
\end{figure}

Without loss of generality, we consider the case where the tip is
laterally displaced from the center of the molecule, e.g., $\vec{a}=\left(0.2\mathrm{nm},0,d\right)$.
We calculate the zero-frequency inelastic current as shown in Fig.
\ref{fig:lateral_displacement} (a). The curve shows the same feature
as that in Fig. \ref{fig:convergency =00003D00003D00003D00003D000026 derivative_1}
(a). The other parameters in Fig. \ref{fig:lateral_displacement}
are the same as that in Fig. \ref{fig:convergency =00003D00003D00003D00003D000026 derivative_1}.
The coincidence between the red dotted line and the blue dashed line
also shows the convergency of our calculation when we consider the
summation with terms $\left|l\right|\leq3$ and $\left|l-l'\right|\leq3$.
We also calculate the first derivative of the zero-frequency inelastic
current with respect to the AC frequency to explore the fine details
of current. As shown in Fig. \ref{fig:lateral_displacement} (b) ,
the first derivative of the zero-frequency inelastic current reveals
the discontinuity at the same spots $\hbar\nu=0.04\mathrm{eV},\;0.06\mathrm{eV},\;\mathrm{and}\;0.12\mathrm{eV}$.
The results in $\vec{a}=\left(0,0,d\right)$ and $\vec{a}=\left(0.2\mathrm{nm},0,d\right)$
indicate that the fine structure in the AC current is robust with
respect to the relative position between the tip and molecule.

\begin{figure}
\begin{centering}
\includegraphics[width=8.5cm]{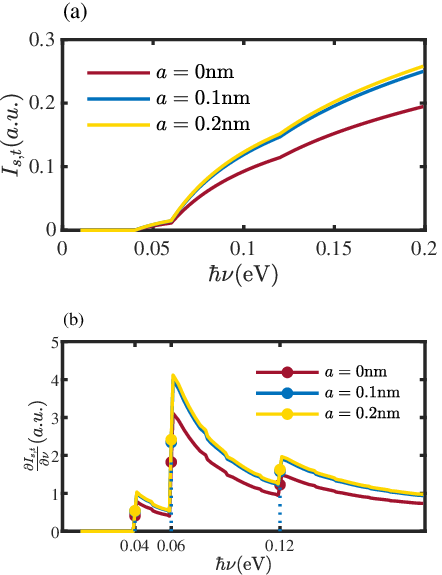}
\par\end{centering}
\caption{\label{fig:different_a}(a) The zero-frequency component of the inelastic
tunneling current as a function of the frequency with the lateral
displacement between the tip and the molecule fixed at $\vec{a}=\left(0,0,d\right)$,
$\vec{a}=\left(0.1\mathrm{nm},0,d\right)$ and $\vec{a}=\left(0.2\mathrm{nm},0,d\right)$,
which is represented by the red, blue and yellow curves respectively.
(b) The first derivative of the zero-frequency component of the current
with respective to the frequency. The correspondence between the color
and the lateral displacement is the same as (a). The dotted markers
show the discontinuous spots in curves. The other parameters are the
same as before.}
\end{figure}

To investigate the influence of the relative position of the tip and
the molecule on the inelastic current, we calculated the zero-frequency
component of the current and its first derivative with respect to
the frequency as a function of the frequency in Fig. \ref{fig:different_a}.
Since we assume that the molecular dipole is isotropic in three axes,
the inelastic current is symmetric around the z-axis. Hence, we only
calculate the displacement of the tip in the x-axis. In Fig. \ref{fig:different_a},
the red, blue and yellow curves correspond to the cases with $\vec{a}=\left(0,0,d\right)$,
$\vec{a}=\left(0.1\mathrm{nm},0,d\right)$ and $\vec{a}=\left(0.2\mathrm{nm},0,d\right)$
respectively. In Fig. \ref{fig:different_a} (a), when the tip is
right above the molecule, the zero-frequency inelastic current is
much smaller than the other cases. The current with $\vec{a}=\left(0.1\mathrm{nm},0,d\right)$
is a little smaller than the current with $\vec{a}=\left(0.2\mathrm{nm},0,d\right)$.
So, the inelastic current can change with the distance between the
tip and the molecule. In Fig. \ref{fig:different_a} (b), the dotted
markers reveal the knee points of the current at same spots $\hbar\nu=0.04\mathrm{eV},\;0.06\mathrm{eV},and\;0.12\mathrm{eV}$,
showing that the resonant relation is independent on the relative
of the tip and the molecule.

\begin{figure*}[t]
\begin{centering}
\includegraphics[scale=0.8]{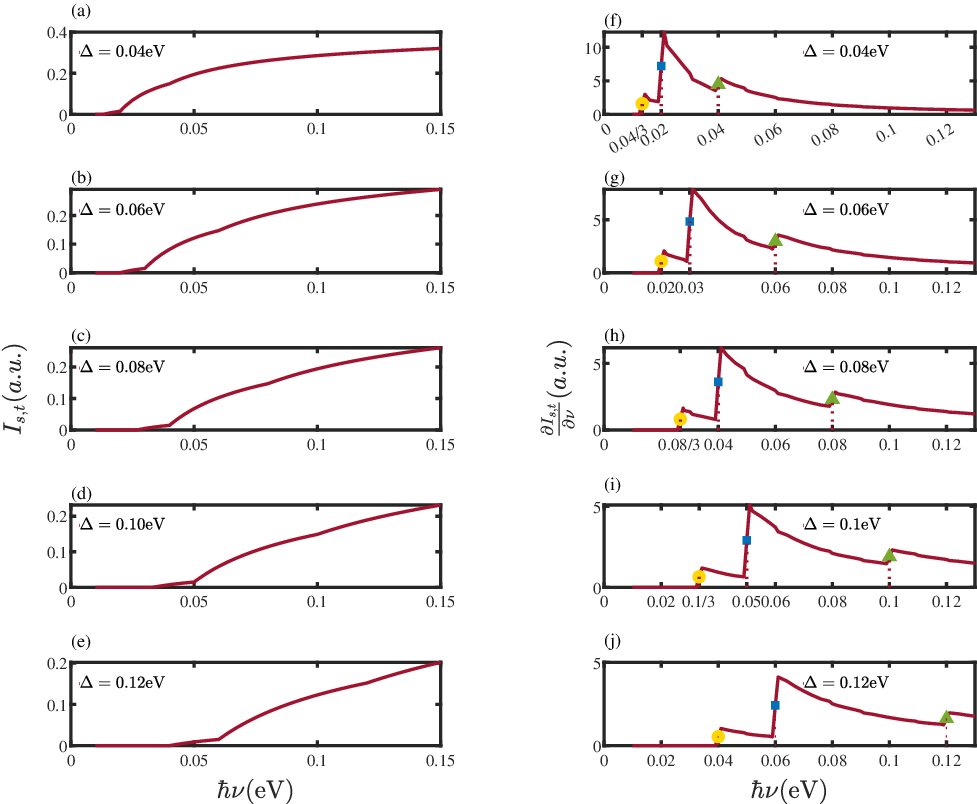} 
\par\end{centering}
\caption{\label{fig:different_delta}The inelastic tunneling current and its
first derivative in $\vec{a}=\left(0.2\mathrm{nm},0,d\right)$ with
$\Delta=0.04,\;0.06,\;0.08,\;0.1,\;\mathrm{and}\;0.12\mathrm{eV}$.
(a)-(e) show the inelastic tunneling current with an upward trend.
(f)-(j) give the first derivative of the current. The round (square,
triangle) dot characterizes the discontinuous points.}
\end{figure*}

To show the general case, we calculate the inelastic tunneling current
and its first derivative with different energy detuning $\Delta=0.04,\;0.06,\;0.08,\;0.1,\;\mathrm{and}\;0.12\mathrm{eV}$,
as illustrated in Fig. \ref{fig:different_delta}. The left column
represents the current curves with an upward trend as discussed before.
When the frequency $\nu$ of the bias is small, the energy of the
tunneling electron is too weak to excite the molecule. No inelastic
tunneling electron transfers energy to the molecule and no inelastic
current flows through the electrodes. The energy of the rising point
in the current becomes higher as $\Delta$ increases. From the definition
$\Delta=E_{\mathrm{eg}}+eV_{0}$, the increase of $V_{0}$ causes
the effective bias $l'\hbar\nu/e-V_{0}$ decreasing. To reach the
molecular excitation energy, the value of $\hbar\nu$ should be larger.
The right column in the Fig. \ref{fig:different_delta} plots the
first derivative of the current. The round (square, triangle) spot
reveals the non-analysis feature of the current. With $\Delta$ increasing,
the energy of the knee point in the same shape increases too.

\begin{figure}[t]
\begin{centering}
\includegraphics{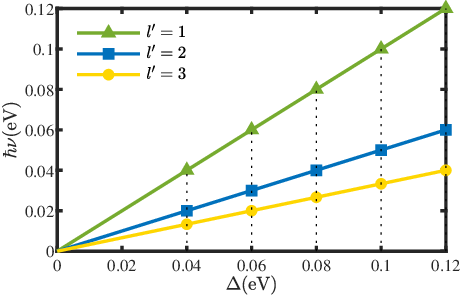}
\par\end{centering}
\caption{\label{fig:point}The energy of knee points as a function of $\Delta=E_{\mathrm{eg}}+eV_{0}$.
The data points in round (square, triangle) correspond to the knee
points with the same marker in Fig. \ref{fig:different_delta}. The
slope of the triangle (square, round) dot line equals to $1/l'=1\left(1/2,1/3\right)$.}
\end{figure}

To figure out the relation between the energy of knee points and $\Delta$,
we show the energy of knee points as the function of the energy $\Delta$
with various orders. Fig. \ref{fig:point} plots the energy of spots
with the same markers in Fig. \ref{fig:different_delta}. The energy
of the knee points linearly increases with energy $\Delta$. The slope
of the line with triangle (square, round) marker is 1 (1/2, 1/3),
and matches $1/l'$ in the relation $\Delta-l'\hbar\nu=0$ with $l'=1\left(2,3\right)$.
This demonstrates that the molecular gap $E_{\mathrm{eg}}$ can be
determined via the knee points $\Delta-l'\hbar\nu=0$.

\section{Conclusions \label{sec:Conclusions}}

In summary, we have proposed the AC-STML setup to measure fine molecular
structures. We calculate the photon counting reflected by the inelastic
current and obtain its Fourier components at the long-time limit.
We show that the rising position of the current spectrum is precisely
determined by the match between the effective bias and the molecular
energy gap. These rising positions are utilized to find the molecular
energy levels by scanning the frequency of the AC bias. The observations
here allow us to propose an alternative method to determine the molecular
levels, especially the fine structures around electronic levels, e.g.,
the vibrational levels.

The AC-STML method can be realized in experiments. Theoretically,
our proposal works well in a large range of AC frequency. In reality,
the AC frequency can be realized around GHz \cite{Gabelli_2008}.
Therefore, as long as we can localize the rough bias $V_{0}$ to the
range $\left|E_{\mathrm{eg}}-e\left|V_{0}\right|\right|\leq10\mu$eV
through the inelastic current curve in the DC bias case, the precise
energy gap will be obtained successfully.

\section*{Acknowledgements}
This work is supported by the National Natural Science Foundation
of China (NSFC) (Grant No. 11875049), the NSAF (Grant Nos. U1730449
and U1930403), and the National Basic Research Program of China (Grant
No. 2016YFA0301201).

\appendix

\section*{Appendix A \label{sec:Appendix}. The derivation of the AC current}

\setcounter{section}{1}

The first derivative of the inelastic tunneling amplitude $c_{e,k}\left(t\right)$
\begin{align*}
\mathrm{i}\hbar\frac{\mathrm{d}c_{e,k}\left(t\right)}{\mathrm{d}t} & =\left(\tilde{\xi}_{k}\left(t\right)+E_{e}\right)c_{e,k}\left(t\right)\\
 & +\mathcal{N}_{s,t}|_{V_{b},E_{n}\rightarrow\xi_{k}}\mathrm{e}^{-\mathrm{i}\left(\tilde{E}_{n}+E_{g}\right)\frac{t}{\hbar}}.
\end{align*}
By using the relation $\widetilde{\xi}_{k}=\xi_{k}+eV_{0}+eV_{1}\sin\left(\nu t\right)$
and $\widetilde{E}_{n}=E_{n}$, we rewrite the inelastic tunneling
amplitude $c_{e,k}\left(t\right)$ as 
\begin{align*}
\frac{\mathrm{d}c_{e,k}\left(t\right)}{\mathrm{d}t} & =\frac{1}{\mathrm{i}\hbar}\left[\xi_{k}+eV_{0}+eV_{1}\sin\left(\nu t\right)+E_{e}\right]c_{e,k}\left(t\right)\\
 & +\frac{1}{\mathrm{i}\hbar}\mathcal{N}_{s,t}|_{V_{b},E_{n}\rightarrow\xi_{k}}\mathrm{e}^{-\mathrm{i}\left(E_{n}+E_{g}\right)\frac{t}{\hbar}}.
\end{align*}
The solution of the inelastic tunneling amplitude $c_{e,k}\left(t\right)$
is 
\begin{align*}
c_{e,k}\left(t\right) & =\frac{1}{\mathrm{i}\hbar}\mathcal{N}_{s,t}|_{V_{b},E_{n}\rightarrow\xi_{k}}\mathrm{e}^{-\mathrm{i}\left[\left(\xi_{k}+E_{e}+eV_{0}\right)\frac{t}{\hbar}-\frac{eV_{1}}{\hbar\nu}\cos\left(\nu t\right)\right]}\\
 & \times\int_{0}^{t}\mathrm{d}\tau\mathrm{e}^{\mathrm{i}\left[\left(\xi_{k}-E_{n}+E_{\mathrm{eg}}+eV_{0}\right)\frac{\tau}{\hbar}-\frac{eV_{1}}{\hbar\nu}\cos\left(\nu\tau\right)\right]}.
\end{align*}
The inelastic tunneling rate can be expressed as 
\begin{align*}
\frac{\mathrm{d}\left|c_{e,k}\left(t\right)\right|^{2}}{\mathrm{d}t} & =\frac{\mathrm{d}}{\mathrm{d}t}\left[c_{e,k}\left(t\right)\cdot c_{e,k}^{*}\left(t\right)\right]\\
 & =\frac{\mathrm{d}c_{e,k}^{*}\left(t\right)}{\mathrm{d}t}\cdot c_{e,k}\left(t\right)+c.c\\
 & =\left\{ -\frac{1}{\mathrm{i}\hbar}\left[\xi_{k}+eV_{0}+eV_{1}\sin\left(\nu t\right)+E_{e}\right]c_{e,k}^{*}\left(t\right)\right.\\
 & \left.-\frac{1}{\mathrm{i}\hbar}\mathcal{N}_{s,t}^{*}|_{V_{b},E_{n}\rightarrow\xi_{k}}\mathrm{e}^{\mathrm{i}\left(E_{n}+E_{g}\right)\frac{t}{\hbar}}\right\} \cdot c_{e,k}\left(t\right)+c.c\\
 & =-\frac{1}{\mathrm{i}\hbar}\mathcal{N}_{s,t}^{*}|_{V_{b},E_{n}\rightarrow\xi_{k}}\mathrm{e}^{\mathrm{i}\left(E_{n}+E_{g}\right)\frac{t}{\hbar}}c_{e,k}\left(t\right)+c.c.
\end{align*}
Substituting the inelastic tunneling amplitude $c_{e,k}\left(t\right)$
into the inelastic tunneling rate $\mathrm{d}\left|c_{e,k}\left(t\right)\right|^{2}/\mathrm{d}t$,
we have 
\begin{align}
\frac{\mathrm{d}\left|c_{e,k}\left(t\right)\right|^{2}}{\mathrm{d}t} & =\frac{1}{\hbar^{2}}\mathcal{N}_{s,t}^{2}|_{V_{b},E_{n}\rightarrow\xi_{k}}\mathrm{e}^{\mathrm{i}\left(E_{n}+E_{g}\right)\frac{t}{\hbar}}\nonumber \\
 & \times\mathrm{e}^{-\mathrm{i}\left[\left(\xi_{k}+E_{e}+eV_{0}\right)\frac{t}{\hbar}-\frac{eV_{1}}{\hbar\nu}\cos\left(\nu t\right)\right]}\nonumber \\
 & \times\int_{0}^{t}\mathrm{d}\tau\mathrm{e}^{\mathrm{i}\left[\left(\xi_{k}-E_{n}+E_{\mathrm{eg}}+eV_{0}\right)\frac{\tau}{\hbar}-\frac{eV_{1}}{\hbar\nu}\cos\left(\nu\tau\right)\right]}+c.c\nonumber \\
 & =\frac{1}{\hbar^{2}}\mathcal{N}_{s,t}^{2}|_{V_{b},E_{n}\rightarrow\xi_{k}}\nonumber \\
 & \times\mathrm{e}^{-\mathrm{i}\left[\left(\xi_{k}+E_{\mathrm{eg}}+eV_{0}-E_{n}\right)\frac{t}{\hbar}-\frac{eV_{1}}{\hbar\nu}\cos\left(\nu t\right)\right]}\nonumber \\
 & \times\int_{0}^{t}\mathrm{d}\tau\mathrm{e}^{\mathrm{i}\left[\left(\xi_{k}-E_{n}+E_{\mathrm{eg}}+eV_{0}\right)\frac{\tau}{\hbar}-\frac{eV_{1}}{\hbar\nu}\cos\left(\nu\tau\right)\right]}+c.c.\label{eq:inelastic_tunneling_rate}
\end{align}

Then Eq. (\ref{eq:inelastic_tunneling_rate})can be rewritten as 
\begin{align*}
\frac{\mathrm{d}\left|c_{e,k}\left(t\right)\right|^{2}}{\mathrm{d}t} & =\frac{1}{\hbar^{2}}\mathcal{N}_{s,t}^{2}|_{V_{b},E_{n}\rightarrow\xi_{k}}\mathrm{e}^{-\mathrm{i}\left[Bt-\frac{eV_{1}}{\hbar\nu}\cos\left(\nu t\right)\right]}\\
 & \times\int_{0}^{t}\mathrm{d}\tau\mathrm{e}^{\mathrm{i}\left[B\tau-\frac{eV_{1}}{\hbar\nu}\cos\left(\nu\tau\right)\right]}+c.c\\
 & =\frac{2}{\hbar^{2}}\mathcal{N}_{s,t}^{2}|_{V_{b},E_{n}\rightarrow\xi_{k}}\mathrm{Re}\left\{ \mathrm{e}^{-\mathrm{i}\left[Bt-\frac{eV_{1}}{\hbar\nu}\cos\left(\nu t\right)\right]}\right.\\
 & \left.\times\int_{0}^{t}\mathrm{d}\tau\mathrm{e}^{\mathrm{i}\left[B\tau-\frac{eV_{1}}{\hbar\nu}\cos\left(\nu\tau\right)\right]}\right\} ,
\end{align*}
where we have defined the notation $B\equiv\left(\xi_{k}-E_{n}+E_{\mathrm{eg}}+eV_{0}\right)/\hbar$.

Using the Jacobi-Anger expansion $\mathrm{e}^{\mathrm{i}a\cos\left(\nu t\right)}=\sum_{l=-\infty}^{\infty}\mathrm{e}^{\mathrm{i}l\pi/2}J_{l}\left(a\right)\mathrm{e}^{\mathrm{i}l\nu t}$
we have 
\begin{align*}
\mathrm{e}^{\mathrm{i}\frac{eV_{1}}{\hbar\nu}\cos\left(\nu t\right)} & =\sum_{l=-\infty}^{\infty}\mathrm{e}^{\mathrm{i}l\pi/2}J_{l}\left(\frac{eV_{1}}{\hbar\nu}\right)\mathrm{e}^{\mathrm{i}l\nu t},\\
\mathrm{e}^{-\mathrm{i}\frac{eV_{1}}{\hbar\nu}\cos\left(\nu t\right)} & =\mathrm{e}^{-\mathrm{i}\frac{eV_{1}}{\hbar\nu}\cos\left(-\nu t\right)}\\
 & =\sum_{l=-\infty}^{\infty}\mathrm{e}^{\mathrm{i}l\pi/2}J_{l}\left(-\frac{eV_{1}}{\hbar\nu}\right)\mathrm{e}^{-\mathrm{i}l\nu t}.
\end{align*}

The inelastic tunneling rate can be expressed through the series of
Bessel's function

\begin{align}
\frac{\mathrm{d}\left|c_{e,k}\left(t\right)\right|^{2}}{\mathrm{d}t} & =\frac{2}{\hbar^{2}}\mathcal{N}_{s,t}^{2}|_{V_{b},E_{n}\rightarrow\xi_{k}}\nonumber \\
 & \sum_{l,l'=-\infty}^{\infty}\mathrm{Re}\left\{ \mathrm{e}^{-\mathrm{i}Bt}\mathrm{e}^{\mathrm{i}l\pi/2}J_{l}\left(\frac{eV_{1}}{\hbar\nu}\right)\mathrm{e}^{\mathrm{i}l\nu t}\right.\nonumber \\
 & \left.\times\int_{0}^{t}\mathrm{d}\tau\mathrm{e}^{\mathrm{i}B\tau}\mathrm{e}^{\mathrm{i}l'\pi/2}J_{l'}\left(-\frac{eV_{1}}{\hbar\nu}\right)\mathrm{e}^{-\mathrm{i}l'\nu\tau}\right\} \nonumber \\
 & =\frac{2}{\hbar^{2}}\mathcal{N}_{s,t}^{2}|_{V_{b},E_{n}\rightarrow\xi_{k}}\nonumber \\
 & \times\sum_{l,l'=-\infty}^{\infty}\left(-1\right)^{l'}J_{l}\left(\frac{eV_{1}}{\hbar\nu}\right)J_{l'}\left(\frac{eV_{1}}{\hbar\nu}\right)\nonumber \\
 & \times\mathrm{Re}\left\{ \mathrm{e}^{-\mathrm{i}\left(B-l\nu\right)t}\mathrm{e}^{\mathrm{i}\left(l+l'\right)\pi/2}\int_{0}^{t}\mathrm{d}\tau\mathrm{e}^{\mathrm{i}\left(B-l'\nu\right)\tau}\right\} ,
\end{align}
where we have used the relation $J_{l}\left(-a\right)=\left(-1\right)^{l}J_{l}\left(a\right)$.
Calculating the integral, we obtain 
\begin{align}
\frac{\mathrm{d}\left|c_{e,k}\left(t\right)\right|^{2}}{\mathrm{d}t} & =\frac{2}{\hbar^{2}}\mathcal{N}_{s,t}^{2}|_{V_{b},E_{n}\rightarrow\xi_{k}}\sum_{l,l'=-\infty}^{\infty}\left(-1\right)^{l'}\nonumber \\
 & \times J_{l}\left(\frac{eV_{1}}{\hbar\nu}\right)J_{l'}\left(\frac{eV_{1}}{\hbar\nu}\right)\left(B-l'\nu\right)^{-1}\nonumber \\
 & \times\mathrm{Re}\left\{ \mathrm{e}^{\mathrm{i}\left[\left(l-l'\right)\nu t+\left(l+l'-1\right)\pi/2\right]}\right.\nonumber \\
 & \left.-\mathrm{e}^{\mathrm{i}\left[-\left(B-l\nu\right)t+\left(l+l'-1\right)\pi/2\right]}\right\} \nonumber \\
 & =\frac{2}{\hbar^{2}}\mathcal{N}_{s,t}^{2}|_{V_{b},E_{n}\rightarrow\xi_{k}}\sum_{l,l'=-\infty}^{\infty}\left(-1\right)^{l'}\nonumber \\
 & \times J_{l}\left(\frac{eV_{1}}{\hbar\nu}\right)J_{l'}\left(\frac{eV_{1}}{\hbar\nu}\right)\left(B-l'\nu\right)^{-1}\nonumber \\
 & \times\left\{ \cos\left[\left(l-l'\right)\nu t+\left(l+l'-1\right)\pi/2\right]\right.\nonumber \\
 & \left.-\cos\left[-\left(B-l\nu\right)t+\left(l+l'-1\right)\pi/2\right]\right\} .\label{eq:Euler bessel inelastic tunneling rate}
\end{align}

The inelastic current is $I_{s,t}\left(t\right)=e\sum_{n,k}\frac{\mathrm{d}}{\mathrm{d}t}\left|c_{e,k}\left(t\right)\right|^{2}$.

By replacing the summation with integral and substituting the inelastic
tunneling rate Eq. (\ref{eq:Euler bessel inelastic tunneling rate})
into the current above $I_{s,t}$$\left(t\right)$, we obtain the
inelastic tunneling current explicitly as (Eq. (\ref{eq:time-depended current})
in the main text) 
\begin{align}
I_{s,t} & \left(t\right)=\int_{-\infty}^{\mu_{0}}\mathrm{d}E_{n}\int_{\mu_{0}}^{0}\mathrm{d}\xi_{k}\rho_{s}\left(E_{n}\right)\rho_{t}\left(\xi_{k}\right)\frac{\mathrm{d}}{\mathrm{d}t}\left|c_{e,k}\left(t\right)\right|^{2}\nonumber \\
 & =\frac{2}{\hbar^{2}}\int_{2\mu_{0}}^{\mu_{0}}\mathrm{d}E_{n}\int_{\mu_{0}}^{0}\mathrm{d}\xi_{k}\rho_{s}\left(E_{n}\right)\rho_{t}\left(\xi_{k}\right)\nonumber \\
 & \times\mathcal{N}_{s,t}^{2}|_{V_{b},E_{n}\rightarrow\xi_{k}}\sum_{l,l'=-\infty}^{\infty}\left(-1\right)^{l'}\nonumber \\
 & \times J_{l}\left(\frac{eV_{1}}{\hbar\nu}\right)J_{l'}\left(\frac{eV_{1}}{\hbar\nu}\right)\left(B-l'\nu\right)^{-1}\nonumber \\
 & \times\{\cos\left[\left(l-l'\right)\nu t+\left(l+l'-1\right)\pi/2\right]\nonumber \\
 & -\cos\left[-\left(B-l\nu\right)t+\left(l+l'-1\right)\pi/2\right]\}.
\end{align}

\vspace*{2mm}

\vspace*{-1mm}
\begin{small}\baselineskip=10pt\itemsep-2pt

 \bibliographystyle{myiopart-num}
\bibliography{iopart-num}

\end{small}
\end{document}